# Large Language Models for Spreadsheets: Benchmarking Progress and Evaluating Performance with FLARE

Simon Thorne, Department of Computer Science, Cardiff Metropolitan University

sthorne@cardiffmet.ac.uk

**ABSTRACT**

*Large Language Models (LLMs) have demonstrated some significant capabilities across various domains; however, their effectiveness in spreadsheet related tasks remains underexplored. This study introduces a foundation for a comprehensive benchmark framework to evaluate the performance of leading LLMs in executing spreadsheet functions, formula generation, and data manipulation tasks. The benchmark encompasses tasks ranging from basic formula creation to complex, real world spreadsheet scenarios. Our findings reveal that while LLMs exhibit proficiency in straightforward tasks, they often falter in complex, multi step operations, frequently producing plausible yet incorrect outputs. These results underscore the limitations of current LLMs in handling spreadsheet tasks that require precise logical reasoning and highlight the need for integrating symbolic reasoning capabilities into LLM architectures. To support this, we introduce FLARE (Formula Logic, Auditing, Reasoning and Evaluation) a new benchmark for evaluating LLM performance on real-world spreadsheet logic, auditing, and reasoning tasks.*

## 1.0 INTRODUCTION

Spreadsheets are integral tools in various sectors, facilitating data analysis, financial modeling, and decision making processes. Despite their ubiquity, spreadsheets are prone to errors, with studies indicating that a significant proportion of operational spreadsheets contain non trivial mistakes (Panko, 2000; Powell, Baker, & Lawson, 2009). These errors can lead to substantial financial and operational consequences.

Large Language Models (LLMs) arrived in late 2022 with ChatGPT3.5, this technology immediately had an impact, providing text and code generation, image creation and stoking a general interest in interacting with Artificial Intelligence (AI) for work and personal interaction. Of particular interest is the ability to generate code from plain English statements, GPT3.5 was able to produce all sorts of code including Excel. As discovered in several papers on the subject published by EuSpRIG, it was noticed that inaccuracies could arise from the process of prompting GPT for spreadsheet code. These inaccuracies are "hallucinations" and are evident not just in code generation but in every task LLMs complete. This was probed by Thorne (2023) and O'Beirne (2023) through a series of tests and experiments designed to explore the ability of GPT3.5 to produce accurate code, statements, logical reasoning and mathematical functions. The findings of this research showed that GPT would regularly hallucinate, had little ability to reason, deduce or infer and was a pure probabilistic system which is simply guessing the "next word" based on the prompt (Jiang et al. 2020).

Research into hallucinations in large language models (LLMs) has intensified as these models are increasingly deployed in real world applications. Hallucinations refer to instances where LLMs generate outputs that are nonsensical, contradict established facts or fail to meet the specification of the input. A comprehensive survey by (Huang, et al., 2025) categorises hallucinations into intrinsic types, arising from the model's internal mechanisms, and extrinsic types, stemming from external factors like ambiguous prompts or insufficient context. Studies have identified specific causes of hallucinations, such as inadequate general knowledge in lower layers of transformer models and issues in the self attention mechanisms of upper layers (Huang, et al., 2025; Verspoor, 2024; Zhang, et al., 2025a; Zhang, et al., 2025b).

**1.1 Motivation**

In 2025, significant change and" progress" has been made in the LLM space, ChatGPT3.5 has evolved into more sophisticated models, OpenAI now has multiple LLM models including "chain of thought" reasoning models, "mini" versions of larger models and some that are designed specifically for coding. OpenAI has been joined by a myriad of other LLM providers such as Microsoft Copilot, Anthropic's Claude AI, DeepSeek, Cohere, Perplexity, Grok3 and many others. All of these models have some ability to generate spreadsheets, spreadsheet formulas, and analysis based on a spreadsheet data. Through regular interaction with these models, it is clear that their ability to generate outputs has evolved significantly since their inception and they are now much more capable of dealing with more complex scenarios. Hence this paper is interested in understanding what progress has been made in terms of the ability of different LLMs to generate reliable spreadsheet formulas, to generate spreadsheets based on written specifications, to generate valid answers to more complex logic and programming requests and whether the level of sophistication achieved by LLMs enables more complex critical activities such as code inspection, error discover and correction. The paper will also consider some standard problems from the spreadsheet literature as a foundation for a LLM spreadsheet benchmark.

**1.2 Research Questions**

1. To what extent has LLM performance improved on real world spreadsheet tasks from GPT 3.5 (2022) to 2025 models across different competency areas?
2. What types of reasoning or debugging behaviours do current LLMs exhibit when faced with flawed or ambiguous spreadsheet code and what are their limitations?
3. Can a modular benchmarking framework, grounded in real world spreadsheet challenges and prior risk research, be constructed to evaluate LLM performance meaningfully over time?

Research questions 1 and 2 will be explored by revisiting some of the testing from Thorne (2023) and augmenting those tests with some additional tests relevant to spreadsheets that cover more complex creation scenarios, difficult formula generation tests for financial modelling, text concatenation and handling, mathematical calculations and logic problems. The tests included in this paper will form the foundation of a spreadsheet LLM testing benchmark.

The general ethos in selecting spreadsheet tasks to benchmark LLMs against is to pick challenging and difficult prompts, the simple reason for this choice is that there seems little point in testing the models against established benchmark questions since these models explicitly train on that data which defeats the objective (Ni, et al., 2025)

**1.3 Related work**

Recent developments in Large Language Models (LLMs) have inspired a new wave of research focused on spreadsheet reasoning, manipulation, and formula generation. A number of benchmark datasets and evaluation frameworks have been proposed to systematically assess the capabilities of LLMs in handling real world spreadsheet tasks.

SpreadsheetBench (Ma, et al., 2024) was introduced as a benchmark for evaluating the formula generation ability of LLMs based on 912 user generated questions from online Excel forums. Unlike synthetic datasets, this benchmark emphasises real world complexity and diversity. Tasks require generating Excel style formulas that generalise across different spreadsheets. Evaluation is conducted using an online judge framework with multiple test cases per instruction to ensure robustness. The study highlights a significant performance gap between human experts and current LLMs, especially in understanding spreadsheet structure and generating correct formulas.

In parallel, SpreadsheetLLM (Dong, et al., 2024) addresses a different challenge: how to represent large and complex spreadsheets efficiently for LLM processing. The authors propose SHEETCOMPRESSOR, an encoding scheme that dramatically reduces token length by leveraging structural anchors, inverted indexing, and data type aware aggregation. The method achieves a 25× compression ratio while improving downstream task accuracy by over 12%. Their "Chain of Spreadsheet" framework further demonstrates improved reasoning in tasks like table detection and classification.

Moving beyond single turn reasoning, SheetAgent (Chen et al., 2025) presents a generalist agent for spreadsheet tasks. The model decomposes tasks into subgoals via a planner, uses a retriever to extract relevant data, and employs iterative reflection to improve results. SheetAgent is evaluated on the newly proposed SheetRM benchmark, designed to reflect complex, ambiguous, and multi step spreadsheet workflows. Results show SheetAgent significantly outperforms baselines, suggesting the utility of structured agent based approaches in long horizon reasoning.

Focusing specifically on reasoning over spreadsheet data, MiMoTable (Li et al. 2025) provides a benchmark of 428 real spreadsheets and 1,719 question answer pairs across domains like finance, education, and architecture. It introduces six meta operation categories: Lookup, Edit, Calculate, Compare, Visualise, and Reasoning each mapped to increasing levels of difficulty. Experimental evaluations reveal a steady performance decline as reasoning complexity increases, with Claude 3.5 Sonnet scoring highest at 77.4% accuracy. The benchmark provides a structured way to test LLMs across multiple reasoning modalities and spreadsheet layouts.

Taken together, these works underscore the rapid evolution in LLM capabilities for spreadsheet tasks from naive formula generation to complex, multi modal reasoning across compressed and multi sheet workbooks. However, as observed in earlier work (Thorne, 2023; O'Beirne, 2023), hallucinations, logical errors, and brittle generalisation remain persistent issues. There remains a clear need for standardised benchmarks and evaluation protocols to assess whether new models can meet the growing demand for accuracy, reliability, and reasoning in spreadsheet based AI systems.

**1.4 Gaps in Current Spreadsheet LLM Benchmarks**

Although there has been clear progress in evaluating the spreadsheet capabilities of LLMs, current benchmarks remain narrow in scope and divorced from the way spreadsheets are actually used. Most focus on isolated formula generation, like translating a prompt into a SUMIF or VLOOKUP, but overlook the cognitive complexity involved in real world spreadsheet work. In practice, users inherit messy, error prone files, diagnose faults, apply domain specific logic, and explain their reasoning to others. This kind of interaction, rich in ambiguity and expectation, is largely absent from existing benchmarks.

A major gap is the lack of tasks involving fault detection or error auditing. Decades of research (Panko, 2008; Baker et al, 2006; Rajalingham et al., 2000; Rakovic et al., 2019; Croll, 2017) has shown spreadsheets are highly error prone, yet LLM benchmarks rarely assess whether a model can identify or explain common issues like named range errors, logical mistakes, or broken conditional formulas. Error seeded tasks, like the Triangle spreadsheet or the Computer Chip Factory are designed to test exactly this, but are missing from current evaluations. Similarly, real world tasks like the Wall and Ball problem (Irons, 2003) show how spreadsheets embed physical principles and assumptions, requiring domain specific reasoning. Benchmarks that ignore this complexity risk overestimating model performance.

In addition, few evaluations probe advanced formula generation, such as corkscrew dynamics, compound interest, BiEntropy measures (Croll, 2013), or word constraint logic (Thorne 2023). These tasks require symbolic abstraction, multi step logic, and generalisation, precisely the capabilities we should be testing in next generation models. Without tasks that span error detection, explanation, domain modelling, and formula reasoning, existing benchmarks remain fragmented. What's needed is a comprehensive, modular framework that reflects real world spreadsheet use and enables meaningful comparison between models over time.

### 1.5 Scope and Positioning of This Work

This paper does not aim to provide a final or exhaustive benchmark for spreadsheet related LLM capabilities. Rather, it introduces an initial framework, a first generation of tasks and evaluation strategies that reflects known challenges in spreadsheet use, logic modelling, and formula reasoning. The benchmark spans areas such as formula generation, error detection, domain modelling, and explanation, but it is not intended to be definitive. Given the complexity and diversity of real world spreadsheet use, it is expected that many of these tasks will evolve over time, with new, more refined test cases emerging to address gaps or improve precision. This work should be seen as a foundation, a structured starting point for evaluating spreadsheet LLM performance designed to be modular and extensible. Future contributions from the research and practitioner community are encouraged to adapt, revise, or augment individual tasks as new capabilities, tools, and use cases emerge.

### 1.6 Methodology

The aim of this study is to evaluate large language model (LLM) performance on tasks that reflect real world spreadsheet usage, including formula generation, error detection, symbolic reasoning, numerical accuracy, and text processing. We propose the FLARE benchmark (*Formula Logic and Reasoning Evaluation*), a modular and extensible test framework specifically designed to reflect the complexity of real-world spreadsheet modelling, error diagnosis, and symbolic computation The benchmark is designed to identify both general purpose reasoning capabilities and domain specific competencies relevant to spreadsheet work in business, finance, engineering, and data analysis.

### 1.7 Selected tasks

The selection of tasks included in this benchmark is an evolution of the tests in Thorne (2023) which reflects a diverse and often underappreciated range of cognitive and functional challenges involved in real world spreadsheet use. The benchmark includes a curated set of problem types that test different dimensions of LLM capability: error detection, symbolic reasoning, domain understanding, numerical abstraction, and explanation. Each task has been chosen for its alignment with known spreadsheet error patterns, its demand for structured logic, or its potential to expose the limitations and emerging capabilities of LLMs. A full list of prompts and responses from the LLMs is available here: https://drive.google.com/drive/folders/1kBGuJoY8dbaIIrnJklo3wghsN4rscYw8?usp=drive_link

#### 1.7.1 Error Seeded "Computer Chip Factory" Spreadsheet
This task replicates common spreadsheet usage by presenting the model with a sheet that contains simple mechanical, logical and omission errors. This is designed to test an LLM's ability to audit and diagnose errors and make corrections. The spreadsheet is a simple profit and loss spreadsheet with some "obvious" seeded errors, such as mechanical data entry errors (inconsistent fixed costs), logical calculation errors (calculating COGS incorrectly) and omission errors (forgetting to include fixed costs in profit calculations).

#### 1.7.2 Error Seeded Triangle spreadsheet
This task provides a more complex auditing task for LLMs, in this case there are a number of error seeded formula using named ranges. This task is "harder" in principle because of the use of abstract named ranges. The formulas for each part of the calculation were extracted and presented to the LLMs in a prompt asking for an accuracy check. There are intentional errors: (1) failure to test whether the sum of the two shorter sides is greater than or equal to the longest side (the triangle inequality condition), (2) returning the wrong variable for the shortest side, (3) returning the wrong variable for the middle side, and (4) incorrectly using the logical operator OR instead of AND when determining the longest side. Each model could score up to eight points, one for detecting and one for correctly fixing each error. This task was presented as a challenge at the 2009 EuSpRIG conference in Paris to the delegates courtesy of Patrick O'Beirne. There are 4 formula errors seeded in this spreadsheet, LLMs will score a point per error they identify and fix.

### 1.7.3 Wall Task
Taken from Panko and Sprague (1998), this task simulates a domain specific cost model involving unit conversions, volume calculation, labour cost, and profit margins. It serves as a test of whether LLMs can build a working spreadsheet from a structured written brief and apply domain appropriate logic. The wall problem was explored in Thorne (2023) with ChatGPT4 and has been used by many researchers as a standard way of evaluating spreadsheet modelling capabilities in humans (Rakovic, et al., 2019; Teo & Tan, 1997; Irons, 2003; Panko & Halverson, 2001; Panko & Sprauge, 1998).

### 1.7.4 The Wall and Ball task
Irons (2003), LLMs to model a safety critical, passenger carrying balloon by integrating geometry, physics, and gas laws into a single, complex scenario. LLMs must compute spherical surface areas and volumes, apply Boyle's Law to adjust for gas compression, and incorporate realistic waste allowances in estimating material and gas usage, ultimately deriving total cost estimates that reflect real world engineering spreadsheets. Given the high stakes of passenger safety, the task is evaluated using stringent, aviation inspired margins of error: outputs within ±0.1% are considered aviation safe, those within ±0.5% are engineering grade, while errors exceeding ±1.5% are unacceptable. The reference benchmark values $266,550.74 for hydrogen and $442,586.96 for helium serve as the "correct" answers against which model performance is measured.

### 1.7.5 Rolling average and Bank Method Interest Calculations
These tasks are included to test LLMs' capacity for handling "difficult" financial modelling domain knowledge, such as dynamic referencing, financial interest conventions, and temporal data modelling. The rolling average task evaluates whether LLMs can generate formulas that correctly calculate a moving average over time, particularly ensuring valid range handling and avoiding out-of-bounds errors. This reflects typical challenges in forecasting, capital modelling, and data smoothing.

The interest calculation task is designed to test domain knowledge, specifically the "Bank Method" of calculating interest, a simplification where all months are assumed to have 30 days, resulting in a 360-day year. This task probes whether LLMs can apply the correct financial basis and distinguish between simple, compound, and amortised interest methods. Success in these tasks would indicate practical readiness for LLM use in financial and economic spreadsheet contexts.

### 1.7.6 BiEntropy Calculations
This task incorporates information theoretic reasoning into spreadsheet calculations (Croll, 2013). BiEntropy offers a challenge that requires not just mathematical accuracy but also interpretation of abstract formulae, requiring the LLM to translate pseudocode or published definitions into valid spreadsheet logic. It is also a test of *adaptability to niche domains*.

### 1.7.8 Latin Square, Word Constraint Logic Puzzles
This category introduces symbolic reasoning tasks such as puzzles requiring constraints on letters, word positions, or logic chains. Though abstract, these problems are ideal for evaluating whether LLMs can apply structured conditional logic, track multiple constraints simultaneously, and reason over symbolic or textual data in spreadsheet form. They also test generalisation, as they are structurally different from traditional numerical problems. An "astronaut specilisation" 5 variable

Latin square puzzle is posed to the LLMs where different astronauts must be assigned to different jobs based on written clues, LLMs will score one mark per correct assignment

**1.8 Scoring**

The LLMs are presented with the test prompts and asked to provide a correct answer in a single shot fashion, i.e. there are no follow up prompts following the initial inquiry. The performance of each test for each LLM is considered in turn, this performance across tests is then bought into a single leaderboard.

The leaderboard ranks the LLMs according to their performance across the six spreadsheet- tasks using a weighted GPA-style scoring system. Each model's performance on a task is scored between 0 and 1, reflecting either full correctness or partial success. For example, a model that correctly identified 3 out of 5 astronaut roles receives a normalised score of 0.6 for that task. These task scores are then combined using the following weighted average formula:

$$\text{Weighted GPA} = (\Sigma(\text{Task Score} \times \text{Task Weight})) / \Sigma(\text{Task Weight})$$

This method ensures that models are evaluated not only on binary correctness but on how much of each task they solve, rewarding partial reasoning and structural effort even when the final result is not perfect.

Each of the six benchmark tasks contributes differently to the final score, based on their complexity and importance in real-world spreadsheet usage. The triangle formula correction task was weighted at 1.5 due to its requirement for symbolic logic repair, forcing models to reason through Boolean expressions and domain rules (e.g., triangle inequality). The bank method interest calculation task, weighted at 1.0, tests whether a model can use the correct financial basis (e.g., a 360-day year) and apply compound or simple interest formulas appropriately.

The text processing task, also weighted at 1.5, assesses a model's ability to manipulate names through string splitting, deduplication, formatting, and alphabetic sorting all in a single Excel formula. The rolling average task, with a weight of 1.0, tests a model's use of relative references and dynamic range logic.

The BiEntropy task was given the highest weight (2.0) because it involved three separate but related symbolic formula generation tasks, each requiring precise logic and numerical reasoning. Finally, the astronaut puzzle, weighted at 1.0, tested deductive reasoning by asking models to assign five individuals to their correct roles based on a set of logical constraints. Though not a formula task, it provided important insight into each model's ability to follow rule-based reasoning and eliminate contradictions.

This scoring methodology balances symbolic logic, functional formula construction, string manipulation, domain-specific knowledge, and logical deduction. It rewards models not just for isolated successes but for breadth and depth of capability across a realistic range of spreadsheet modelling challenges. The weighted GPA system ensures that the resulting leaderboard reflects the true functional spreadsheet competence of each LLM.

**2.0 RESULTS**

The following sections explore the results obtained from various LLMs on the problems posed. The bank of LLMs used in these tests are: Perplexity, Mistral, Cohere, DeepSeek (and R1), GROK3 (and variations), Gemini Flash (and variations), Claude, Copilot, MetaAI, ChatGPT4o (and variations).

## 2.1 The Wall task

| LLM | Final Bid (Welsh Granite) | Final Bid (Brick) | Accuracy |
|---|---|---|---|
| Perplexity, Claude, Cohere, GPT 4o, GPT 4o mini, GPT 4.5, Copilot, GROK3, GROK3 DeepSearch, DeepSeek, DeepSeekR1, Gemini DeepResearch, Gemini 2.5 Flash, Gemini 2.5 Pro, GPTo3 mini, GPTo3 mini high | £1934.40 | £1622.40 | Accurate |
| Mistral | £2683.20 | £2371.20 | Inaccurate |
| GPT 4 legacy | £2683.20 | £2371.20 | Inaccurate |
| Meta | £2310 | £1996.80 | Inaccurate |
| Gemini 2.0 Flash | £2308.80 | £1996.80 | Inaccurate |

Table 1 Wall task results

Table 1 shows the outcome of the Wall task for all LLMs, as can be seen 16 of the models get this correct but 5 models come to the wrong answer.

## 2.2 The Wall and the Ball

| Error Band | LLMs |
|---|---|
| Aviation Grade (≤ 0.1%) | ChatGPT 3 mini high, ChatGPTo1 |
| Suitable (0.1%–0.5%) | Mistral, Gemini Legacy, Perplexity, GROK3, DeepSeek R1, ChatGPT 4o, ChatGPT 4.5, ChatGPT 3.5, ChatGPT 3 mini, Gemini Flash Thinking, GROK3 Deep Search, Copilot, Gemini 2.5 Pro, Gemini DeepResearch |
| Preliminary Use (0.5%–1%) | Claude, Gemini Flash, DeepSeek |
| Borderline (1%–1.5%) | ChatGPT 4 Legacy |
| Unsuitable (> 1.5%) | Cohere |

*Table 2 LLM performance in Wall and Ball task*

Table 2 Wall & Ball Error Bands

The results for the Wall and the Ball are categorised according to the level of error present in the calculations, see table 2. In this case, margin of error are considered because some LLMs calculate aspects of the problem to differing degrees of accuracy based on how many decimal places certain parameters are calculated to and different approaches to rounding values, for instance . The context here is a person carrying balloon and as such demands the highest standards of accuracy to ensure the proposed designs are safe enough to implement.

## 2.3 Error seeded computer chip factory task

| LLMs | Errors Detected | Total Errors | Accuracy (%) |
|---|---|---|---|
| Gemini 2.5 Pro | 12 | 12 | 100.0 |
| ChatGPT4.5, ChatGPTo3 mini high | 11 | 12 | 91.67 |
| ChatGPT4 Legacy, ClaudeAI, GROK3 DeepSearch, MetaAI | 10 | 12 | 83.33 |
| ChatGPT4o, ChatGPTo3 mini, GROK3, Mistral | 9 | 12 | 75.0 |
| ChatGPTo1, DeepSeek, DeepSeekR1 | 8 | 12 | 66.67 |
| Cohere, Perplexity | 7 | 12 | 58.33 |

| Copilot, Gemini Flash Thinking | 4 | 12 | 33.33 |
| Gemini Flash | 3 | 12 | 25.0 |
| ChatGPT4o mini | 2 | 12 | 16.67 |

**Table 3 LLM performance in the "computer chip" simple error seeded spreadsheet**

Across the evaluation of 20 LLMs on a spreadsheet containing 12 seeded errors, we observed clear patterns in the types of mistakes that models were most likely to miss, see table 3. Logical errors, such as incorrect formula references in the January and February rows, were generally well detected by advanced models like ChatGPT4.5, Gemini 2.5 Pro, and GROK3 DeepSearch. In contrast, mechanical errors proved more challenging, many models missed obvious numeric anomalies (e.g. a unit price of 50 instead of 500), spelling mistakes like "Decemeber," and typos in column headers (e.g. "Distrubtion"). Omission errors, such as incomplete formulas that left out fixed costs in profit calculations, were inconsistently detected, with several models overlooking subtle issues in the November rows.

Models like Gemini Flash, Copilot, and Cohere consistently missed multiple mechanical and omission based errors, while Gemini 2.5 Pro and ChatGPT4.5 demonstrated high accuracy across all error types. This suggests that while LLMs are becoming increasingly adept at identifying formula logic flaws, they remain vulnerable to surface level data issues unless specifically prompted to perform checks for consistency, spelling, or outlier values.

## 2.4 Error Seeded Triangle spreadsheet

| LLMs | Group | Error Handling Summary | Score |
|---|---|---|---|
| DeepSeekR1, Perplexity, ChatGPT4.5, ChatGPTo3 mini, ChatGPTo3 mini high, Mistral, Perplexity | A Missed all errors | No detections or fixes | 0 |
| DeepSeek, Cohere, ClaudeAI, ChatGPT4o, ChatGPT4o mini, ChatGPTo3 mini | B Partial detection | Detected triangle inequality only | 1 |
| MetaAI | C Best performer | Fixed triangle inequality and OR→AND | 3 |
| Gemini Flash Thinking, Gemini 2.5 Pro, ChatGPT4o legacy, ChatGPTo1, GROK3, GROK3 DeepSearch | D Partial fix only | Fixed OR→AND only (not detected) – achieved via min, max and median | 1 |
| Gemini Flash | E Partial detect + fix | Detected triangle inequality; fixed OR→AND | 2 |

**Table 4 LLM performance in the triangle error seeded task**

Table 4 shows the results of the Triangle formula error seeded spreadsheet where the LLMs were asked to detect and correct four seeded logic errors in triangle classification formulas.

The majority of models (Group A) failed to detect or fix any of the four errors, indicating a basic lack of formula comprehension or verification ability. A slightly more capable cluster (Group B) correctly identified the missing triangle inequality check but did not propose a valid fix. One standout model, MetaAI, both detected and correctly fixed the triangle inequality issue and also corrected the OR/AND logical error, making it the highest scoring model overall (3 out of 8). A further group (Group D) proposed correct fixes for the OR→AND issue without explicitly detecting it, suggesting partial internal consistency or pattern correction but not clear reasoning. Gemini Flash (Group E) was the only other model to detect the triangle inequality and also fix the OR→AND error, though it failed to apply a proper correction to the inequality itself.

Importantly, none of the models detected or fixed the subtle but crucial errors involving misassignment of variables in the shortest and middle side functions. This suggests that such errors involving symbolic reasoning or reference remain challenging for even the best performing LLMs. While several models showed promising partial capabilities, no LLM approached comprehensive understanding or correction across all four errors.

**2.5 Bank method interest**

| LLMs | Method Used | Correct? |
| --- | --- | --- |
| GROK3, Cohere, Copilot, DeepSeek, Gemini 2.5 Flash | Simple interest, APR ÷ 12 | Incorrect |
| Perplexity, Gemini 2.0 Flash | Amortised interest using IPMT | Incorrect |
| Claude | (P × APR × Days)/360 using 31 days | Correct |
| DeepSeekR1 | Interest with 366-day year | Incorrect |
| META | 365-day basis with 31 days | Incorrect |
| Mistral | Complex IF formula with days per month | Incorrect |
| ChatGPT o4-mini-high | DAYS360 with date ranges | Correct |
| ChatGPT o4-mini | DAYS360 with SEQUENCE | Correct |
| ChatGPT 4.5 | Simple interest × 10 months | Incorrect |
| ChatGPT 4o mini | PMT for full payments | Incorrect |
| ChatGPT 4 Legacy | Compound interest with FV | Incorrect |
| Gemini 2.5 Pro | LET with DaysInMonth and 360-day base | Correct |
| GROK3 DeepSearch | 365-day year with 31 days | Incorrect |

Table 5 LLM performance in the bank method interest test

The evaluation of responses from various large language models (LLMs) to the task of calculating monthly bank method interest revealed that only a few models accurately applied the correct formula. The bank method, which calculates simple interest using the actual number of days in each month divided by a 360 day year, was correctly implemented by Claude, ChatGPT o4 mini high, ChatGPT o4 mini, and Gemini 2.5 Pro. These models used formulas that accounted for monthly day counts and the 360 day convention, either through direct calculations or dynamic functions like DAYS360. In contrast, most other models such as GROK3, Perplexity, DeepSeek, and various versions of ChatGPT and Gemini misapplied simplified interest calculations, compound interest formulas, or amortised methods (e.g., using IPMT or PMT), which are inconsistent with the bank method. This suggests that while many LLMs can produce plausible formulas, only a few reliably apply domain specific conventions accurately without explicit prompting.

**2.6 Text processing**

| LLM | Method Summary | Correct? |
|---|---|---|
| Copilot | Combines cells, splits by comma, dedupes, sorts, joins with TEXTJOIN. Does not reformat to 'Lastname, Firstname'. | No |
| Cohere | Attempts string manipulation with MID, SUBSTITUTE, SORT, and JOIN. Output unclear; likely malformatted. | No |
| Claude | Uses FILTERXML to extract elements, but output retains original 'Firstname Lastname' format. | No |
| Perplexity | Uses FILTERXML and TEXTJOIN to split, dedupe, and join names; output matches correct format. | Yes |
| Gemini | Attempts LET and TEXTSPLIT to extract names and reformat, but output is incorrect and malformed. | No |
| Mistral | Splits and sorts names but keeps 'Firstname Lastname' format and includes duplicates. | No |
| GROK3 | Fails due to syntax error in LET – 'too few arguments'. | No |
| DeepSeek | Splits and dedupes names but keeps them as 'Firstname Lastname' and includes duplicate due to case mismatch. | No |
| DeepSeek R1 | Returns a formula error – 'too few arguments'. | No |
| ChatGPT 4o | Produces a #NAME? error. | No |
| ChatGPT 01 | Syntax error – 'problem with this formula'. | No |
| ChatGPT 03 mini | Invalid formula – results in Excel error. | No |
| ChatGPT 03 mini high | LET block invalid – first argument is not a valid name. | No |
| ChatGPT 4o mini | Formula error – invalid use of TEXTBEFORE/TEXTAFTER. | No |
| ChatGPT 4 Legacy | Produces a formula error from malformed XML string. | No |
| ChatGPT 04 mini | Uses LET, MAP, TEXTSPLIT, SORT, and TEXTJOIN to split, dedupe, reformat as 'Lastname, Firstname', and recombine. | Yes |
| ChatGPT 04 mini high | Uses TEXTSPLIT, PROPER, UNIQUE, SORT, TEXTBEFORE, and TEXTAFTER to generate correct format with correct sorting and casing. | Yes |

Table 6 LLM performance in the text processing task

Table 6 evaluates the performance of various LLMs when asked to generate a nested Excel formula that transforms two cells of delimited names into a single cell with names deduplicated, alphabetically sorted, and reformatted as "Lastname, Firstname", separated by semicolons. Out of the models tested, only Perplexity, ChatGPT-04 mini, and ChatGPT-04 mini high produced correct results that fully satisfied the criteria. A few others, such as Copilot and DeepSeek, came close but failed due to formatting issues or case-sensitive duplicates. The majority of models either produced syntax errors, failed to reformat the names correctly, or attempted overly complex constructions that did not execute in Excel. Notably, many of the GPT-4 derivatives and mini variants failed to handle Excel's functional constraints, particularly with LET, MAP, and TEXTSPLIT.

## 2.7 Rolling Average

| LLMs | Method Used | Correct? |
|---|---|---|
| GROK3, GROK3 DeepSearch, Perplexity, Copilot, META, Gemini Flash 2.5, Gemini 2.5 Pro, ChatGPT4.5, ChatGPT4 Legacy, Mistral | `=AVERAGE(A1:C1)` dragged across or from early cell | No |
| Gemini Flash, ChatGPT o3, ChatGPT o4 mini, ChatGPT o4 mini high, ChatGPT 4o mini, DeepSeek / DeepSeekR1, Gemini 2.5 Pro, Cohere | `OFFSET()` with `IF` or dynamic functions like `MAP`, `MAKEARRAY`, `TRANSPOSE` | Yes |

Table 7 LLM performance in the Rolling Average task

This evaluation groups language models by the type of Excel formula they suggested for calculating a 3 value rolling average across A1:J1. Models that relied solely on dragging `=AVERAGE(A1:C1)` are marked incorrect due to risk of referencing out of bounds cells. A 'Mixed' category captures models like Meta AI that gave both a correct `OFFSET` based formula and an incorrect `AVERAGE` drag variant. Correct models exclusively used safe, bounded formulas using `OFFSET`, `IF`, or dynamic array functions such as `MAP` or `MAKEARRAY`, ensuring valid results without manual adjustments.

## 2.8 BiEntropy calculations

The evaluation of BiEntropy calculations across 16 large language models (LLMs) reveals substantial variability in both accuracy and consistency, see table 7. Only a small subset of models Gemini 2.4 Pro, GROK3 DeepSearch, ChatGPT o3, and Cohere produced results within 0.01 of the correct values for all three binary strings (10011101, 10011001, and 10100101). While several models, including ChatGPT4o, ChatGPT o4 mini high, and DeepSeekR1, successfully calculated the BiEntropy for 10011101, they tended to overestimate the BiEntropy of the more structured strings, especially 10011001.

| LLM | 10011101 | 10011001 | 10100101 | 10011101 Correct? | 10011001 Correct? | 10100101 Correct? |
|---|---|---|---|---|---|---|
| ChatGPT4o | 0.9540 | 0.7520 | 0.8300 | True | False | False |
| Perplexity | 0.9510 | 0.9510 | 0.1070 | True | False | True |
| Copilot | 0.4735 | 0.0116 | 0.0534 | False | False | False |
| Gemini 2.4 Pro | 0.9506 | 0.0234 | 0.1073 | True | True | True |
| META | 0.9506 | 0.4714 | 0.4552 | True | False | False |
| GROK3 DeepSearch | 0.9350 | 0.0230 | 0.1070 | False | True | True |
| ChatGPT4.5 | 0.9502 | 0.7492 | 0.8266 | True | False | False |
| DeepSeek | 0.9460 | 0.9550 | 0.9710 | True | False | False |
| ChatGPT o3 | 0.9510 | 0.0234 | 0.1073 | True | True | True |
| ChatGPT o4 mini high | 0.9540 | 0.9850 | 0.9180 | True | False | False |
| Cohere | 0.9506 | 0.0234 | 0.1073 | True | True | True |
| Gemini 2.5 Flash | 0.8500 | 0.8800 | 1.0000 | False | False | False |
| Gemini 2.0 | 0.9540 | 0.7520 | 0.8290 | True | False | False |
| ChatGPT Legacy | 0.4730 | 0.0120 | 0.0530 | False | True | False |

| Mistral | 0.9532 | 0.1980 | 0.2163 | True | False | False |
| DeepSeekR1 | 0.9547 | 0.7525 | 0.8385 | True | False | False |

**Table 8 LLM performance in the BiEntropy calculation task**

### 2.9 Latin square word puzzle

| LLM | Stellar | Orion | Nebula | Ace | Ryder | Correct? |
|---|---|---|---|---|---|---|
| GROK3 | Astro | Geology | Navigation | Comms | Robotics | No |
| Perplexity | Astro | Geology | Navigation | Comms | Robotics | No |
| Claude | Astro | Geology | Navigation | Comms | Robotics | No |
| Cohere | Robotics | Astro | Geology | Comms | Navigation | No |
| Gemini 2.5 Pro | Astro | Geology | Navigation | Comms | Robotics | No |
| Gemini 2.0 Flash | Astro | Geology | Navigation | Comms | Robotics | No |
| DeepSeek | Astro | Geology | Navigation | Comms | Robotics | No |
| DeepSeekR1 | Astro | Geology | Navigation | Comms | Robotics | No |
| Meta AI | Astro | Geology | Navigation | Comms | Robotics | No |
| Mistral | Comms | Astro | Geology | Robotics | Navigation | No |
| ChatGPT 4 mini | Geology | Astro | Navigation | Comms | Robotics | No |
| ChatGPT 4o | Astro | Geology | Navigation | Comms | Robotics | No |
| ChatGPT 4o mini high | Astro | Geology | Navigation | Comms | Robotics | No |
| ChatGPT 4 Legacy | Comms | Astro | Geology | Robotics | Navigation | No |
| Copilot | Astro | Geology | Navigation | Comms | Robotics | No |
| GROK3 DeepSearch | Astro | Geology | Navigation | Comms | Robotics | No |

**Table 9 LLM performance in the Latin Square test**

The results from this comparative analysis of LLM responses to the astronaut specialisation puzzle reveal an interesting limitation in current model reasoning capabilities. Despite the problem being logically solvable with a unique correct answer, none of the models tested including leading systems like GPT 4, Claude, Gemini, and GROK were able to arrive at the fully correct solution. Most models succeeded in satisfying individual clues in isolation but failed to reconcile all constraints simultaneously. These errors highlight the models' difficulty with multi variable constraint satisfaction problems, particularly when solutions require propagating exclusions across dependent assignments.

Structurally, the puzzle functions as a constrained Latin square logic problem a type of grid based puzzle in which each variable must appear exactly once per row and column, subject to additional logical conditions. Notably, several models provided detailed, confident reasoning chains that sounded plausible but contained subtle logical flaws. This underscores a core limitation: LLMs can imitate reasoning but struggle to systematically deduce the logic needed for Latin square problems. LLMs make predictions based on statistical relationships, not deterministic logic.

### 2.10 Overall scoring

The table below presents the final weighted scores for all LLMs across the benchmark, reflecting performance across five spreadsheet competency areas.

| Model | Triangle | Bank Method | Text Proc | Rolling Avg | BiEntropy | Astronaut | Weighted GPA |
|---|---|---|---|---|---|---|---|
| Gemini 2.5 Pro | 0.125 | 1.000 | 0.500 | 1.000 | 1.000 | 0.400 | 0.667 |
| ChatGPT-4o mini-high | 0.125 | 1.000 | 1.000 | 1.000 | 0.330 | 0.400 | 0.593 |
| ChatGPT 4o | 0.125 | 0.000 | 0.500 | 1.000 | 1.000 | 0.400 | 0.542 |

| Model | | | | | | | |
|---|---|---|---|---|---|---|---|
| Cohere | 0.125 | 0.000 | 0.500 | 1.000 | 1.000 | 0.400 | 0.542 |
| ChatGPT-o3 mini | 0.000 | 0.000 | 0.500 | 1.000 | 1.000 | 0.200 | 0.494 |
| ChatGPT-o3 mini-high | 0.000 | 0.000 | 0.500 | 1.000 | 1.000 | 0.200 | 0.494 |
| Perplexity | 0.125 | 0.000 | 1.000 | 0.500 | 0.660 | 0.400 | 0.488 |
| GROK3 DeepSearch | 0.125 | 0.500 | 0.500 | 0.500 | 0.660 | 0.400 | 0.457 |
| Meta | 0.375 | 0.500 | 0.250 | 1.000 | 0.330 | 0.400 | 0.437 |
| Claude | 0.125 | 1.000 | 0.500 | 0.500 | 0.330 | 0.400 | 0.437 |
| ChatGPT-o4 mini | 0.125 | 1.000 | 0.500 | 1.000 | 0.000 | 0.400 | 0.417 |
| Gemini 2.0 Flash | 0.250 | 0.000 | 0.500 | 1.000 | 0.330 | 0.400 | 0.398 |
| DeepSeek | 0.125 | 0.500 | 0.250 | 1.000 | 0.330 | 0.400 | 0.390 |
| GROK3 | 0.125 | 0.500 | 0.000 | 0.500 | 0.660 | 0.400 | 0.363 |
| Gemini 2.5 Flash | 0.125 | 0.500 | 1.000 | 0.000 | 0.000 | 0.400 | 0.323 |
| DeepSeekR1 | 0.000 | 0.500 | 0.000 | 1.000 | 0.330 | 0.400 | 0.320 |
| Mistral | 0.000 | 0.000 | 0.000 | 0.500 | 0.330 | 0.400 | 0.195 |
| Copilot | 0.000 | 0.250 | 0.250 | 0.500 | 0.000 | 0.400 | 0.191 |
| ChatGPT 4-mini | 0.125 | 0.000 | 0.000 | 0.000 | 0.000 | 0.400 | 0.073 |

**Table 10 LLM performance leaderboard**

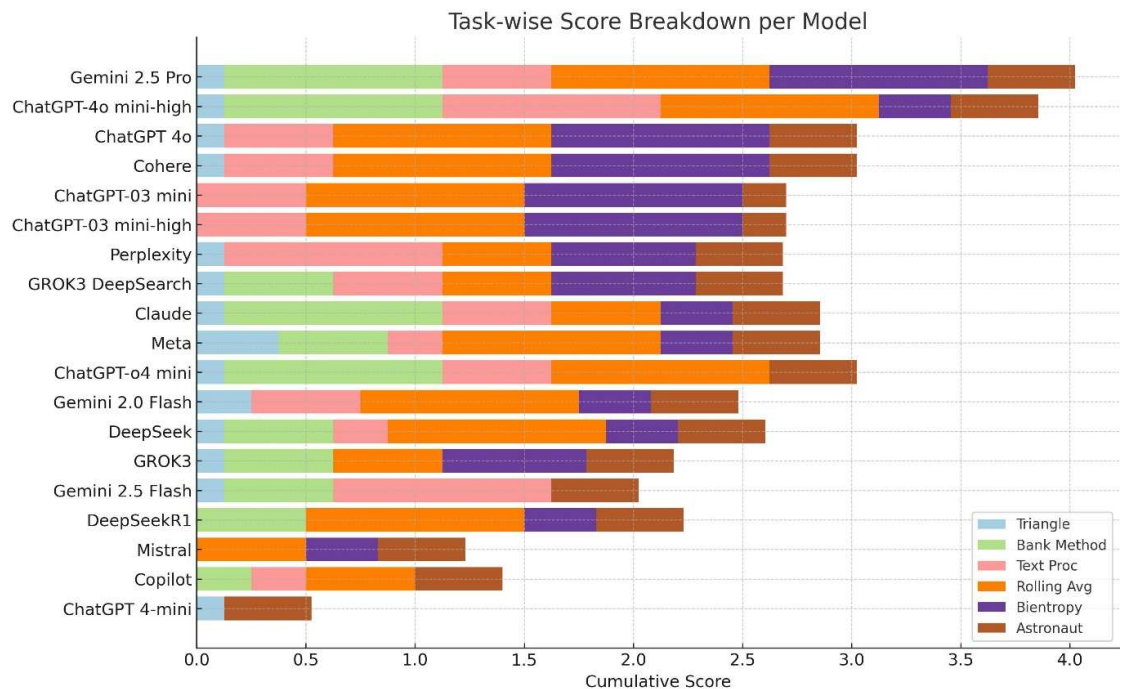

**Figure 1 LLM performance across tasks**

Table 10 and Figure 1 show the final leaderboard and paints a clear and nuanced picture of current LLM capabilities in spreadsheet reasoning. Gemini 2.5 Pro leads the field with a Weighted GPA of 0.636, showing some consistency across all six tasks, particularly in symbolic reasoning (BiEntropy), financial modelling (bank method), and structured text transformation. Close behind, ChatGPT-4o mini-high and ChatGPT-4o mini show strength in applied logic and formula generation, though the latter doesn't perform as well in symbolic and text-heavy tasks. Claude, Cohere, and DeepSeekR1 cluster in the middle tier, performing capably in isolated domains, particularly symbolic logic or relative referencing but falter in multi-step formatting or domain-specific scenarios. At the lower end, models like ChatGPTo4 mini, Copilot and Mistral consistently underperform, particularly on tasks requiring precise text handling or formula repair, reaffirming the gap between newer, tuned architectures and legacy models.

When considering the performance by task, the highest average scores were observed in rolling average and bank method tasks, formulaic scenarios where models could match learned templates to well-structured inputs. This suggests that modern LLMs are increasingly adept at replicating familiar patterns, especially when the problem is framed procedurally. In contrast, symbolic reasoning tasks like the BiEntropy and triangle spreadsheet, produced wide variation and lower average scores, with many models failing to identify logic errors or generalise formula structures. The astronaut logic puzzle proved challenging across the board although some models achieved partial matches, none could resolve all five constraints correctly, highlighting their probabilistic and deterministic logic nature. Text processing, particularly name formatting and reordering, showed the lowest average scores overall, underlining the brittleness of LLM outputs when deterministic string control is required.

Considering this benchmark against recent spreadsheet benchmarks such as SpreadsheetBench (Ma et al., 2024), SheetAgent (Chen et al., 2025), and MiMoTable (Li et al., 2025) represent valuable advances in evaluating formula generation, multi-modal reasoning, and agent-based task planning, they remain limited in scope or realism. SpreadsheetBench focuses primarily on formula generation from natural language prompts, but lacks auditing, symbolic reasoning, or domain-specific modelling tasks. MiMoTable introduces real spreadsheets and reasoning categories, but does not evaluate formula correction or fault diagnosis. SheetAgent takes an agent-based approach to task decomposition but abstracts away the low-level spreadsheet structure and does not assess formula correctness at the cell level. In contrast, the benchmark proposed in this paper is designed to test LLM performance in real-world, error-prone, domain-specific spreadsheet tasks, including symbolic logic, text transformation, and financial reasoning. By grounding each task in well-established spreadsheet risk scenarios, this framework complements existing benchmarks while addressing their limitations, particularly around logic repair, fault explanation, and deterministic reasoning in spreadsheet contexts.

## 3.0 SUMMARY

The following sections close off the paper, they consider the research questions set, pose the question has there been real progress with LLMs, consider the limitations and future work and finally conclude the paper.

### 3.1 Revisiting the Research Questions

This next section will consider the original research questions posed at the start of this paper.

> RQ1. To what extent has LLM performance improved on real-world spreadsheet tasks from GPT-3.5 (2022) to 2025 models across different competency areas?

The leaderboard confirms measurable improvement in LLM spreadsheet performance since 2022, but also clarifies the limits of this progress. While GPTo4, Copliot, Mistral and similar legacy models struggled with even basic spreadsheet logic, newer entrants such as Gemini 2.5 Pro and ChatGPT-4o mini-high now demonstrate some competence across several domains, especially structured domain modelling and formula generation. Performance is not uniformly distributed most models exhibit siloed ability, working well in template-based tasks but failing at symbolic abstraction or logic puzzle resolution. Overall, the benchmark reveals clear generational gains but also exposes the fragility of these improvements when applied to more complex or less familiar problems, i.e. problems that are significantly different to the types of problems commonly found in regular LLM benchmarks.

> RQ2. What types of reasoning or debugging behaviours do current LLMs exhibit when faced with flawed or ambiguous spreadsheet code and what are their limitations?

LLMs show superficial but not deep debugging capabilities. In tasks like the triangle spreadsheet, many models offered syntactically correct but logically inappropriate rewrites frequently changing function structure without actually resolving the seeded error. In some cases, models identified that something was wrong but failed to apply domain logic to fix it. More advanced models (e.g., Gemini 2.5 Pro, Claude) showed signs of intermediate reasoning, but still struggled with multi-step dependencies or contradictions within the problem space. A recurring limitation is that most models fail silently: they produce plausible outputs that conceal logical flaws and offer no self-checking. While some debugging behaviour is evident, this remains context-dependent and inconsistent. The results suggest current LLMs cannot yet be trusted with spreadsheet debugging without close human oversight but could potentially be useful as collaborators.

> RQ3. Can a modular benchmarking framework, grounded in real-world spreadsheet challenges and prior risk research, be constructed to evaluate LLM performance meaningfully over time?

Yes. The benchmark framework developed in this study has successfully captured performance variance across model generations and task types. It draws on known spreadsheet risk scenarios and includes tasks that span numeric modelling, symbolic logic, and error detection. Low scores across the board validate the difficulty and relevance of the tasks rather than undermining the benchmark's utility. As such, this framework serves as a reliable foundation for longitudinal assessment and model comparison.

### 3.2 Progress?

While progress in LLM capability is evident, especially in structured modelling and common formula generation, the overall scores remain low with even the best models achieving less than two-thirds of available points. This suggests that improvements have been narrow rather than deep. Higher performing models like Gemini 2.5 Pro and ChatGPT-4o mini-high succeed when the task resembles training data or has a template-like structure but continue to struggle with symbolic reasoning, constraint satisfaction, and compositional logic. The benchmark highlights the nature of LLMs, probabilistic guessing based on statistical inference. Their outputs often appear fluent and plausible, but close inspection reveals brittle logic, hidden flaws, or hallucinations particularly in tasks requiring careful logic (e.g., astronaut puzzle, BiEntropy).

The appearance of fluency may obscure serious errors, particularly in mission critical spreadsheet applications. Confident errors are more dangerous than obvious failures, and most models lack any mechanism to indicate uncertainty or validate their outputs. As such, they may give users a false sense of reliability. These findings call into question the trustworthiness of LLMs in unsupervised spreadsheet use. Without advances in logical robustness, model introspection, and error explanation, LLMs should be treated as assistive tools requiring human oversight not autonomous agents.

### 3.3 Usability, Trust and the Nature of LLM knowledge

While this study has primarily focused on benchmarking and comparative evaluation, an important underlying question remains: *Are these models actually usable in contexts where mistakes are unacceptable?* In most real-world spreadsheet applications, particularly in finance, engineering, and regulatory environments, even small errors can have serious consequences. What makes the use of LLMs uniquely risky is the *confidence* with which they present outputs, often cloaked in a patina of technical plausibility.

This observation invites a broader reflection on the epistemological status of LLM-generated knowledge. Unlike human reasoning, which draws on context, intention, and responsibility, LLMs operate as probabilistic engines, selecting tokens based on frequency patterns rather than grounded understanding. This raises a critical sociotechnical question: *What kind of knowledge is this?*

LLMs may *resemble* intelligence in their outputs, but they lack awareness, intentionality, and the capacity for accountability. In this light, the benchmark results do not just chart functional capabilities—they also reveal the fragility and impersonality of machine-generated certainty. As these systems become embedded in workflows, there is a danger that fluent, confident outputs will obscure their limitations, especially for non-expert users. This creates a paradox: the better LLMs become at mimicking competence, the harder it becomes to detect when they are fundamentally guessing.

In this sense, future work must not only improve LLM performance but also grapple with their ontological and ethical position in knowledge production. A deeper integration of LLM benchmarking with the philosophy and sociology of knowledge will be essential for evaluating when, and *if* LLMs should be trusted in high-stakes domains.

### 3.4 Limitations and Future Work

This study remains constrained by several factors. First, although grounded in real-world spreadsheet risks, the benchmark only samples a limited set of task types. Areas such as dynamic dashboarding, VBA integration, and interactive workflows were not assessed. Second, while the scoring rubric balances granularity and practicality, partial credit always introduces interpretive subjectivity, even under a rubric-based scheme.

Third, models were tested using static prompts. This does not reflect interactive LLM use, where clarifying prompts or step-wise decomposition might improve performance. Prompt chaining, agent frameworks, or user-guided iteration are important future directions. Finally, the study evaluated only the text-in/text-out behaviour of models; future work should explore multimodal spreadsheets, hybrid systems, and LLMs embedded within existing spreadsheet environments.

### 3.5 CONCLUSION

This paper presents a comprehensive evaluation of LLMs against a spreadsheet benchmark grounded in real-world modelling, error detection, and symbolic logic tasks. The results confirm that while LLM performance has improved markedly since GPT-3.5, especially in structured and templated domains, the underlying reasoning remains brittle, shallow, and fragile across logic-intensive tasks. The benchmark reveals both meaningful gains and serious limitations. The current generation of LLMs even at their best lack the consistency, introspection, and logic fidelity required for autonomous spreadsheet generation or auditing. The FLARE benchmark provides a structured and discriminative tool for tracking progress in LLM capabilities related to spreadsheet tasks, offering a more realistic and rigorous alternative to synthetic or narrow benchmarks.

The benchmark itself proves effective, modular, and extensible, offering a realistic and discriminative way to track LLM development over time. It highlights the importance of symbolic and structural logic areas that remain unsolved and require deeper innovation. Until LLMs can reason, self-check, and explain, they must be deployed as fallible assistants requiring human experts to validate output. Future progress must focus not just on performance metrics, but on model reliability, transparency, and trustworthiness in spreadsheet-intensive domains.